\documentclass{article}
\usepackage[utf8]{inputenc}  
\usepackage{geometry} 
\usepackage[colorlinks = true, linkcolor = red, urlcolor = red, citecolor = red, anchorcolor = red]{hyperref}
\usepackage{graphicx}  
\usepackage{datetime}  
\newdate{date}{21}{11}{2023}
\usepackage{amsmath,amssymb,amsfonts,amsthm}  
\usepackage{algorithm}
\usepackage{algorithmic}
\usepackage{mathtools}
\usepackage[usestackEOL]{stackengine}
\stackMath
\usepackage{caption}
\usepackage{newfloat}
\usepackage{listings}
\DeclareCaptionStyle{ruled}{labelfont=normalfont,labelsep=colon,strut=off} 
\floatstyle{ruled}
\newfloat{listing}{tb}{lst}{}
\floatname{listing}{Listing}

\newcommand{\qrlew}{\emph{Qrlew}}

\title{\qrlew: Rewriting SQL into Differentially Private SQL}
\author {
    Nicolas Grislain\\
    \href{mailto:ng@sarus.tech}{ng@sarus.tech}
    \and
    Paul Roussel\\
    \href{mailto:pr@sarus.tech}{pr@sarus.tech}
    \and
    Victoria de Sainte Agathe\\
    \href{mailto:vdsa@sarus.tech}{vdsa@sarus.tech}
}
\date{\displaydate{date}}

\begin{document}

\maketitle

\begin{abstract}
This paper introduces \qrlew{}, an \emph{open source} library that can parse SQL queries into \emph{Relations} --- an intermediate representation --- that keeps track of rich data types, value ranges, and row ownership; so that they can easily be rewritten into \emph{differentially-private} equivalent and turned back into SQL queries for execution in a variety of standard data stores.

With \qrlew{}, a \emph{data practitioner} can express their data queries in standard SQL; the \emph{data owner} can run the rewritten query without any technical integration and with strong privacy guarantees on the output; and the query rewriting can be operated by a privacy-expert who must be trusted by the owner, but may belong to a separate organization.
\end{abstract}

\section{Introduction}

In recent years, the importance of safeguarding privacy when dealing with personal data has continuously increased.
Traditional anonymization techniques have proven vulnerable to re-identification, as demonstrated by numerous works \cite{archie2018s, dwork2017exposed, narayanan2008robust, sweeney2013identifying}.
The total cost of data breaches has also significantly increased \cite{ibm2023cost} and governments have introduced stricter data protection laws.
Yet, the collection, sharing, and utilization of data holds the potential to generate significant value across various industries, including healthcare, finance, transportation, and energy distribution.

To realize these benefits while managing privacy risks, researchers have turned to \emph{differential privacy (DP)} \cite{wood2018differential, dwork2014algorithmic}, which has become the \emph{gold standard} for privacy protection since its introduction by Dwork et al. in 2006 \cite{dwork2006calibrating} due to its provable and automatic privacy guarantees.

Despite the availability of powerful open-source tools \cite{kotsogiannis2019privatesql, diffprivlib, OpenDP, PipelineDP, ZetaSQL, PrivacyOnBeam, johnson2020chorus, berghel2022tumult, yousefpour2021opacus}, DP adoption remained limited and many organizations sticked to more manual and \emph{ad-hoc} approaches.
Reasons for this lack of adoption are probably complex and multiple but one could name: the lack of awareness on privacy risks; the loss of utility in the results; and the perceived complexity of the existing solutions considering they all require, either some expertise in differential privacy, or the use of new interfaces to express data processing tasks, or even to integrate new execution engines in their data stack.
\qrlew{} \cite{Grislain_Qrlew_2023} has been designed to relieve these problems by providing the following features:
\begin{description}
    \item[\qrlew{} provides automatic output privacy guarantees]
    With \qrlew{} a \emph{data owner} can let an analyst (\emph{data practitioner}) with no expertise in privacy protection run arbitrary SQL queries with strong privacy garantees on the output.
    \item[\qrlew{} leverages existing infrastructures]
    \qrlew{} rewrites a SQL query into a \emph{differentially private} SQL query that can be run on any data-store with a SQL interface: from lightweight DB to big-data stores.
This removes the need for a custom execution engine and enables \emph{differentially private analytics with virtually no technical integration}.
    \item[\qrlew{} leverages synthetic data]
    Synthetic data are an increasingly popular way of \emph{privatizing} a dataset \cite{bowen2019comparative, mckenna2021winning, canale2022generative, sablayrolles2023privately, castellon2023dp}. Using jointly: \emph{differentially private} mechanisms and \emph{differentially private} synthetic data can be a simple, yet powerful, way of managing a privacy budget and reaching better utility-privacy tradeoffs.
\end{description}


\section{Definitions}

\subsection*{Datasets and Privacy Units (PU)}

In this paper, \emph{datasets} refer to a collection of elements in some domain $\mathcal{X}$, labelled with an identifier $i\in \mathcal{I}$ identifying the entity whose privacy we want to protect. This entity will be called \emph{Privacy Unit} (PU) and the identifier will be referred to as \emph{Privacy ID} (PID). Let $\mathcal{D}$ be the set of datasets of arbitrary sizes with a privacy unit.

\subsection*{Differential Privacy (DP)}

Let $\mathcal{M}$ be an algorithm that takes a dataset as input and produces a randomized output. The algorithm $\mathcal{M}$ is said to satisfy $\varepsilon,\delta$-differential privacy if, for all pairs of adjacent datasets $D, D' \in \mathcal{D}$, and for all measurable sets $S$ in the range of $\mathcal{M}$:
$$\Pr[\mathcal{M}(D) \in S] \leq e^{\varepsilon} \cdot \Pr[\mathcal{M}(D') \in S] + \delta$$

\subsection*{Adjacent datasets}

Datasets $D, D' \in \mathcal{D}$ are adjacent if they are equal up to the addition or removal of all entries sharing the same PID. Note that this is a slightly unusual and restricted definition of adjacency, suited to our practical needs. It is close to that used in the \emph{user-level differential privacy} literature \cite{liu2020learning, wilson2019differentially} where one user can have many samples.

\section{Assumptions and Design Goals}

In this work, we assume the \emph{central model of differential privacy} \cite{near2020threat}, where a trusted central organization: hospital, insurance company, utility provider, called the \emph{data owner}, collects and stores personal data in a secure database and whishes to let untrusted \emph{data practitioners} run SQL queries on its data.

At a high level we pursued the following requirements:
\begin{itemize}
    \item Ease of use for the \emph{data practitioners}. The \emph{data practitioners} are assumed to be a data experts but no privacy experts. They should be able to express their queries in a standard way. We chose SQL as the query language as it is very commonly used for analytics tasks.
    \item Ease of integration for the \emph{data owner}. As SQL is a common language to express data analysis tasks, many data-stores support it from small embedded databases to big data stores.
    \item Simplicity for the \emph{data owner} to setup privacy protection. Differential privacy is about capping the sensitivity of a result to the addition or removal of an individual that we call \emph{privacy unit}. \qrlew{} assumes that the \emph{data owner} can tell if a table is public and, if it is not, that it can assign exactly one \emph{privacy unit} to each row of data. In the case there are multiple related tables, \qrlew{} enables to define easily the \emph{privacy units} for each tables transitively.
    \item Simple integration with other privacy enhancing technologies such as \emph{synthetic data}. To avoid repeated privacy losses or give result when a DP rewriting is not easily available (e.g. when the query is: \texttt{SELECT * FROM table}) \qrlew{} can use \emph{synthetic data} to blend in the computation.
\end{itemize}

These requirements dictated the overall \emph{query rewriting} architecture and many features, the most important of which, are detailed below.

\section{Architecture and main features of \qrlew}

\begin{figure*}[t]
    \centering
    \includegraphics[width=\textwidth]{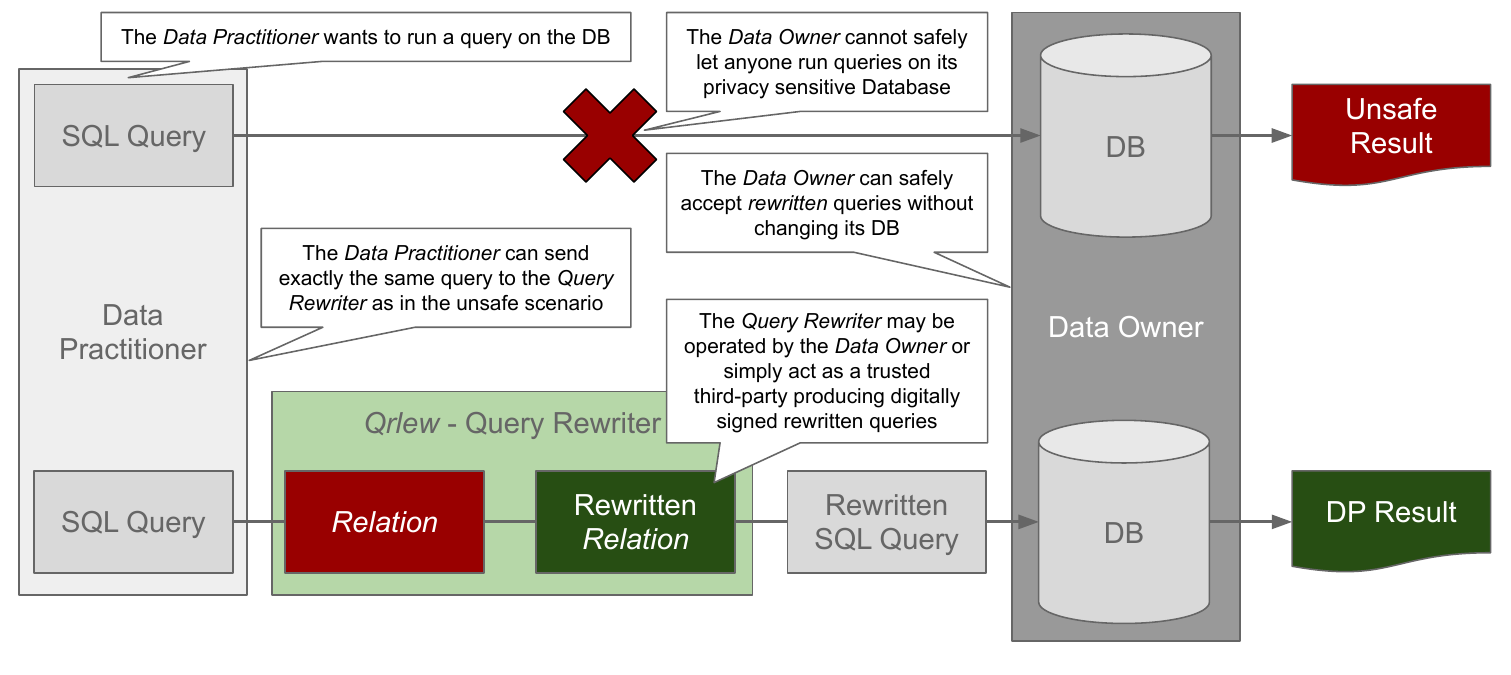} 
    \caption{The rewriting process occurs in three stages: The \emph{data practitioner}'s query is parsed into a \emph{Relation}, which is rewritten into a DP equivalent and finally executed by the the \emph{data owner} which returns the privacy-safe result.}
    \label{fig:process}
\end{figure*}

The \qrlew{} library, solves the problem of running a SQL query with DP guarantees in three steps. First the SQL query submitted by the \emph{data practitioner} is parsed and converted into a \emph{Relation}, this \emph{Relation} is an intermediate representation that is designed to ease the tracking of data types ranges or possible values, to ease the tracking of the \emph{privacy unit} and to ease the rewriting into a DP \emph{Relation}. Then, the rewriting into DP happens. Once the relation is rewritten into a DP one, it can be rendered as an SQL query string and submitted to the data store of the \emph{data owner}. The output can then safely be shared with the \emph{data practitioner}. This process is illustrated in figure~\ref{fig:process}.

\subsection{Qrlew Intermediate Representation}
\label{sec:intermediate_representation}

As the SQL language is very rich and complex, simply parsing a query into an abstract syntax tree does not produce a convenient representation for our needs. Therefore, it is converted into a simpler normalized representation with properties well aligned with the requirements of Differential Privacy: the \emph{Relation}. A \emph{Relation} is a collection of rows adhering to a given \emph{schema}. It is a recursively defined structure composed of:
\begin{description}
    \item[\emph{Tables}] This is simply a data source from a database.
    \item[\emph{Maps}] A \emph{Map} takes an input \emph{Relation}, filters the rows and transform them one by one. The filtering conditions and row transforms are expressed with expressions similar to those of SQL. It acts as a \texttt{SELECT exprs FROM input WHERE expr LIMIT value} and therefore preserve the \emph{privacy unit} ownership structure.
    \item[\emph{Reduces}] A \emph{Reduce} takes an input \emph{Relation} and aggregates some columns, possibly group by group. It acts as a \texttt{SELECT aggregates FROM input GROUP BY expr}. This is where the rewriting into DP will happen as described in section~\ref{sec:rewriting}.
    \item[\emph{Joins}] This \emph{Relation} combines two input \emph{Relations} as a \texttt{SELECT * FROM left JOIN right ON expr} would do it. The privacy properties are more complex to propagate in this case.
\end{description}
It may also be a static list of values or a set operation between two \emph{Relations}, but those are less important for our uses.

\begin{figure}[t]
    \centering
    \includegraphics[width=0.7\textwidth]{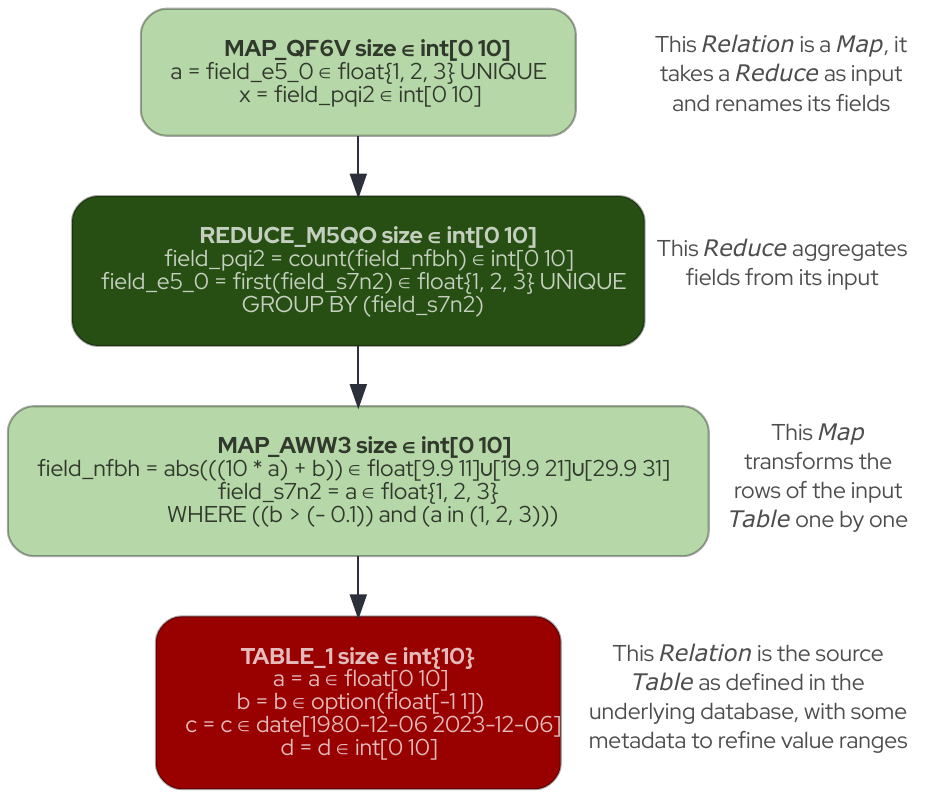}
    \caption{\emph{Relation} (\emph{Map}) associated to the query: \texttt{SELECT a, count(abs(10*a+b)) AS x FROM table\_1 WHERE b>-0.1 AND a IN (1,2,3) GROUP BY a}. The arrows point to the inputs of each \emph{Relation}. Note the propagation of the data type ranges.}
    \label{process}
\end{figure}

This representation is central to \qrlew{}; all the features described below are built upon it. A \emph{Relation}, along with all the sub-\emph{Relations} it depends on, will be called the \emph{computation graph} or the \emph{graph} of a \emph{Relation}.

\subsection{Range Propagation}
\label{sec:range_propagation}

Most DP mechanisms aggregating numbers require the knowledge of some bounds on the values (see \cite{dwork2014algorithmic}).
Even if some bounds are known for some \emph{Relations} like source \emph{Tables}, it is not trivial to propagate these bounds through the steps of the computation.

To help with range propagation, \qrlew{} introduces two useful concepts:
\begin{itemize}
    \item The concept of \emph{$k$-Interval}, which are finite unions of at most $k$ closed intervals. A \emph{$k$-Interval} can be noted:
    $$I = \bigcup_{i=1}^{j\leq k}\left[a_i, b_i\right]$$
    Note that the union of \emph{$k$-Interval}s may not be a \emph{$k$-Interval} as it may be the union of more than $k$ intervals.
    Unions of many intervals can be simplified into their convex envelope interval, which are often sufficient bounds approximations for our use cases:
    $$J = \bigcup_{i=1}^{j> k}\left[a_i, b_i\right] \subseteq \left[\min_i a_i, \max_i b_i\right]$$
    \item And the concept of \emph{piecewise-monotonic-functions}\footnote{Which is a shorthand name for what would be better called: \emph{piecewise-coordinatewise-monotonic-functions}}, which are functions $f: \mathbb{R}^n \rightarrow \mathbb{R}$ whose domain can be partitioned in cartesian products of intervals: $P_j$ on which they are \emph{coordinatewise-monotonic}.
    The image of a cartesian product of $n$ \emph{$k$-Interval}s by a \emph{piecewise-monotonic-function} can be easily computed as a \emph{$k$-Interval}.
    Indeed, let $I$ be:
    $$I = I_1\times I_2\times \ldots \times I_n = \bigcup_{\substack{1\leq i_1\leq k\\\ldots\\1\leq i_n\leq k}}\left[a_{i_1}, b_{i_1}\right]\times \ldots \times \left[a_{i_n}, b_{i_n}\right]$$
    If $f$ is \emph{piecewise-monotonic}, then one can show that on each partition $P_j$ where it is \emph{coordinatewise-monotonic}, if we note:
    $$I_j = I \cap P_j = \bigcup_{\substack{1\leq j_1\leq k\\\ldots\\1\leq j_n\leq k}}\left[a_{j_1}, b_{j_1}\right]\times \ldots \times \left[a_{j_n}, b_{j_n}\right]$$
    $$f(I_j) = \bigcup_{\substack{1\leq i_1\leq k\\\ldots\\1\leq i_n\leq k}}\text{Conv}\left(f\left( \left\{a_{i_1}, b_{i_1}\right\}\times \ldots \times \left\{a_{i_n}, b_{i_n}\right\}\right)\right)$$
    where $\text{Conv}\left(f\left( \left\{a_{i_1}, b_{i_1}\right\}\times \ldots \times \left\{a_{i_n}, b_{i_n}\right\}\right)\right)$ can be efficiently computed in $n$ steps, without testing all the $2^n$ combinations, thanks to the coordinatewise monotony of $f$ on $P_j$.
    Then $f(I) = \bigcup_j f(I_j)$, of which we can derive the bounding: $f(I) \subseteq \text{Conv}\left(\bigcup_j f(I_j)\right)$ when the number of terms in the union exceeds $k$.
\end{itemize}

The notion of \emph{$k$-Interval} is convenient for tracking value bounds as it can express natural patterns in SQL such as:
\begin{itemize}
    \item \texttt{WHERE x>0 AND x<=1}, which translates into the implied $x\in \left[0, 1\right]$ ;
    \item \texttt{WHERE x IN (1,2,3)}, which is also easily expressed as a \emph{k-Interval}: $x \in \left[1, 1\right] \cup \left[2, 2\right] \cup \left[3, 3\right]$
\end{itemize}

The idea of \emph{piecewise-monotonic-function} is also very useful as in SQL many standard arithmetic operators (\texttt{+}, \texttt{-}, \texttt{*}, \texttt{/}, \texttt{<}, \texttt{>}, \texttt{=}, \texttt{!=}, \ldots) and functions (\texttt{EXP}, \texttt{LOG}, \texttt{ABS}, \texttt{SIN}, \texttt{COS}, \texttt{LEAST}, \texttt{GREATEST}, \ldots) are trivially \emph{piecewise-monotonic-function} (in one, two or many variables).

Most of the range propagation in \qrlew{} is based on these concepts. It enables a rather simple and efficient range propagation mechanism, leading to better utility / privacy tradeoffs.

\subsection{Privacy Unit Definition}
\label{sec:privacy_unit_definition}

Tables in a database rarely come properly formatted for privacy-preserving applications. Many rows in many tables may refer to the same individual, hence, \emph{adding or removing an individual} means \emph{adding or removing many rows}. To help the definition of the privacy unit \qrlew{} introduces a small Privacy Unit (PU) description language.
As exemplified in listing~\ref{lst:pe}, PU definition associates to each private table in a database a path defining the PID of each row. For a table containing the PU itself, like a \texttt{users} table for example, the PU definition will look like \texttt{("users",[],"id"),} where \texttt{id} is the name of a column identifying the user, like its name. If the database defines tables related to this tables, the way the tables are related should be specified following this scheme: $(\mathtt{tab}_1, path, \mathtt{pid})$ where $\mathtt{tab}_1$ is the name of the table for which the PID is defined, $\mathtt{pid}$ is the name of the column defining the PID in the table referred by $path$ and $path$ is a list of elements of the form $[(\mathtt{ref}_1, \mathtt{tab}_2, \mathtt{id}_2),\ldots, (\mathtt{ref}_{m-1}, \mathtt{tab}_m, \mathtt{id}_m)]$
where $\mathtt{ref}_{i-1}$ is a column in $\mathtt{tab}_{i-1}$ --- usually a foreign key --- referring to $\mathtt{tab}_i$ with a column of referred id $\mathtt{id}_i$ --- usually a primary key. Following the path of tables referring to one another, we end up with the table defining the PID (e.g. \texttt{users}).

This small PU description language allows for a variety of useful PID scenarii, beyond the simple, but restrictive \emph{privacy per row}.

\begin{listing}[tb]
\caption{Example of \emph{privacy unit} definition for a database with three tables holding users, orders and items records. Each user is protected individually by designating their \texttt{id}s as PID. Orders are attached to a user through the foreign key: \texttt{user\_id}. Items's ownership is defined the same way by specifying the lineage: \texttt{item -> order -> user}.}%
\label{lst:pe}
\begin{lstlisting}[language=Python]
privacy_unit = [
    ("users",[],"id"),
    ("orders",[
    ("user_id","users","id")
    ],"id"),
    ("items",[
    ("order_id","orders","id"),
    ("user_id", "users", "id")
    ],"id")
]
\end{lstlisting}
\end{listing}

\subsection{Rewriting}
\label{sec:rewriting}

\begin{figure*}[t]
    \centering
    \includegraphics[width=\textwidth]{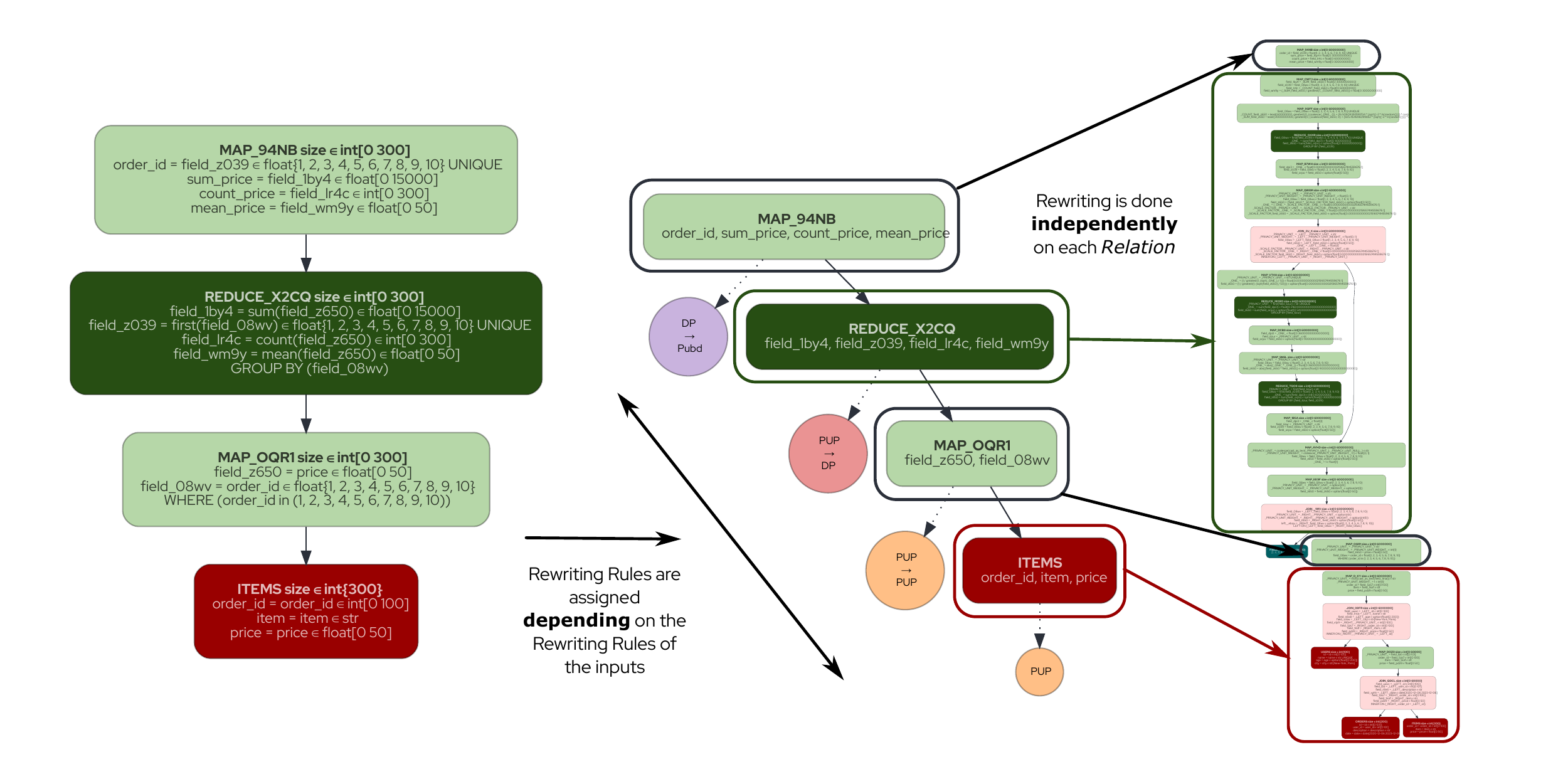} 
    \caption{The rewriting process happens in two phases: a \emph{rewriting rule allocation} phase, where each node in the \emph{computation graph} gets allocated a \emph{rewriting rule} (RR) compatible with its input and with the desired output property; and a \emph{rule application} phase, where each \emph{Relation} is rewritten according to its allocated RR.}
    \label{fig:rewriting}
\end{figure*}

Rewriting in \qrlew{}, refers to the process of altering the \emph{computation graph} by substituting computation \emph{sub-graphs} to \emph{Relations} (see figure~\ref{fig:rewriting}) to alter the properties of the result. This substitution aims to achieve specific objectives, such as ensuring privacy through the incorporation of differentially private mechanisms. The rewriting process (see figure~\ref{fig:rewriting}) happens in two phases:
\begin{itemize}
    \item a \emph{rewriting rule allocation} phase, where each \emph{Relation} in the \emph{computation graph} gets allocated a \emph{rewriting rule} (RR) compatible with its input and with the desired output property;
    \item a \emph{rule application} phase, where each \emph{Relation} is rewritten to a small \emph{computation graph} implementing the logic of the rewriting and stitched together with the other rewritten \emph{Relations}.
\end{itemize}

Before we decribe these phases into more details, let's define the various properties we may want to guarantee on each \emph{Relation} and the ones we need for the output.

\subsubsection{Privacy Properties and Rewriting Rules}
\label{sec:privacy_properties}

Each \emph{Relation} can have one of the following properties:
\begin{description}
    \item[Privacy Unit Preserving (PUP)]: A \emph{Relation} is PUP if each row is associated with a PU. In practice it will have a column containing the PID identifying the PU.
    \item[Differentially Private (DP)] A \emph{Relation} will be DP if it implements a DP mechanism. A DP \emph{Relation} can be safely  executed on private data and the result be published. Note that the \emph{privacy loss} associated with the DP mechanism has to be accurately accounted for (see section~\ref{sec:privacy_analysis}).
    \item[Synthetic Data (SD)] In some contexts a \emph{synthetic data} version of source tables is available. Any \emph{Relation} derived from other SD or Public \emph{Relations} is itself SD.
    \item[Public (Pub)] A relation derived from public tables is labeled as such and does not require any further protection to be disclosed.
    \item[Published (Pubd)] A relation is considered Published if its input relations are either Public, DP, in some cases SD, or Published themselves. It can be considered as Published but with some more care like the need to account for the privacy loss incurred by its DP ancestors.
\end{description}

These properties usually require some rewriting of the computation graph to be achieved. The requirements for a specific \emph{Relation} to meet some property are embodied in what we call: \emph{rewriting rules}.
A \emph{rewriting rule} has input requirements, and an achievable output \emph{property} that tells what \emph{property} can be achieved by rewriting provided the input \emph{property} requirements are fulfilled.
Each \emph{Relation} can be assigned different \emph{rewriting rules} depending on their nature: \emph{Map}, \emph{Reduce}, etc. and the way they are parametrized.

\emph{Rewriting Rules} can be --- for instance --- PU propagation rules of the form:
\begin{itemize}
    \item $\varnothing \rightarrow PUP$ for private \emph{Tables} with a simple rewriting consisting in taking the definition of the privacy unit and computing the PID column.
    \item $PUP \rightarrow PUP$ for \emph{Maps} (or for \emph{Reduce} when the PID is in the \texttt{GROUP BY} part) with a rewriting consisting in propagating the PID column from the input to the output.
    \item $(PUP, PUP) \rightarrow PUP$ (or its variants with one published input) for \emph{Join} and a rewriting consisting in adding the PID in the \texttt{ON} clause.
\end{itemize}

Another key \emph{Rewriting Rules} is $PUP \rightarrow DP$ for \emph{Reduces}, it simply means that if the parent of the \emph{Relation} can be rewritten as PUP, then we can rewrite the relation to be DP by substituting DP aggregations to the original aggregations of the \emph{Reduce}.

One easily see that by simply applying $PUP \rightarrow PUP$ and $PUP \rightarrow DP$ rules, one can propagate the privacy unit across the computation graph of a \emph{Relation} and compute some DP aggregate such as a noisy sum or average.

\subsubsection{Rewriting Rule Allocation}

\begin{figure*}[t]
    \centering
    \includegraphics[width=0.9\textwidth]{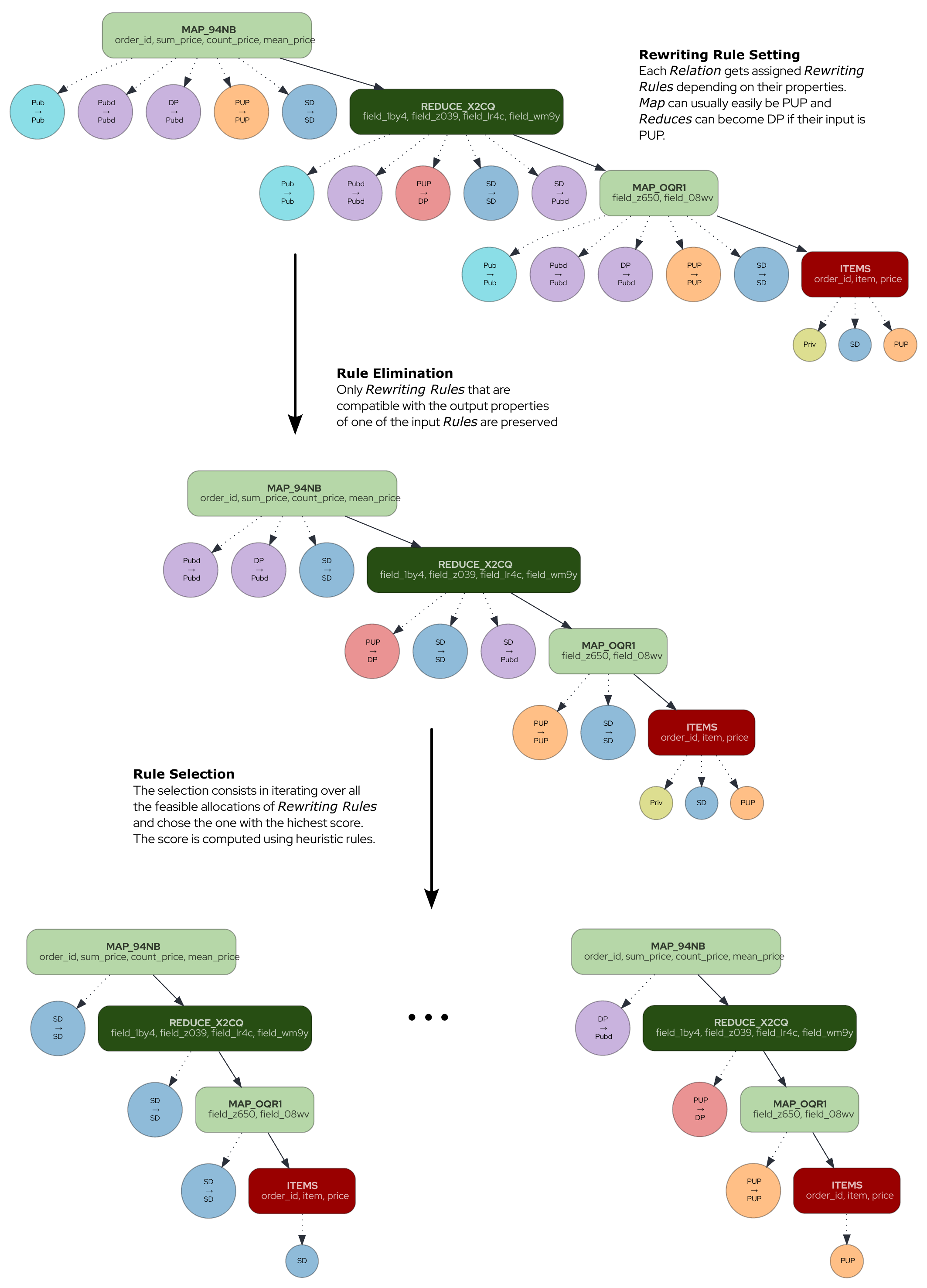} 
    \caption{The rewriting happens in three steps: \emph{Rule Setting} when we assign the set of potential rewriting rules to each \emph{Relation} in a computation graph; \emph{Rule Elimination}, when only feasible rewriting rules are preserved; and \emph{Rule Selection}, when an actual allocation is selected.}
    \label{fig:set_eliminate_select}
\end{figure*}

The first phase of the rewriting process consists in allocating one and only one rule to each \emph{Relation}.
This is done in three steps illustrated in figure~\ref{fig:set_eliminate_select}:
\begin{description}
    \item[Rule Setting] We assign the set of potential rewriting rules to each \emph{Relation} in a computation graph.
    \item[Rule Elimination] Only feasible rewriting rules are preserved. A rewriting rule that would require a PUP input is only feasible if its input Relation has a feasible rule outputting a PUP \emph{Relation}.
    \item[Rule Selection] All feasible allocations of one rewriting rule per \emph{Relation} are listed, a score depending on the desired ultimate output property is assigned to each allocation and the highest scoring allocation is selected. Then, a simple split $\left(\frac{\varepsilon}{n}, \frac{\delta}{n}\right)$ of the overall privacy budget $\left(\varepsilon, \delta\right)$ depending on the number of $PUP \rightarrow DP$ rules: $n$ is chosen.
\end{description}

In the computation graph, while each node's multiple rewriting rules might suggest a combinatorial explosion in the number of possible feasible allocations, this is mitigated in practice. The pruning of infeasible rules, dictated by the requirement for most relations to have a PUP input for a DP or PUP outcome, significantly reduces the complexity. Hence, despite the theoretical breadth of possibilities, the actual number of feasible paths remains manageable, avoiding substantial computational problems in practice.

\subsubsection{Rule Application}

Once the first phase of rule allocation is achieved, starts the second phase: \emph{rule application}, as illustrated in figure~\ref{fig:rewriting}.
In the allocation phase, a \emph{global rewriting scheme} was set in the form of an allocation satisfying a system of requirements; in the rewriting phase, each rewriting rule is applied \emph{independently} for each \emph{Relation}. This is possible because once a rewriting rule is applied to a \emph{Relation}, the \emph{Relation} is transformed into a computation graph of \emph{Relations} whose ultimate inputs are compatible (same schema, i.e. same columns with same types, plus the new columns provided by the property achieved) with the inputs of the original \emph{Relation} and the ultimate output is also compatible with the output of the original \emph{Relation} so that rewritten \emph{Relations} can be stitched together in a larger graph the same way the original \emph{Relations} were connected: see figure~\ref{fig:rewriting}.

\section{Privacy Analysis}
\label{sec:privacy_analysis}

When rewriting, a user can require the output \emph{Relation} to have the \emph{Published} property. All \emph{rewriting rules} with \emph{Published} outputs require their inputs to be either \emph{Public}, \emph{DP}, \emph{SD} or \emph{Published} themselves. We assume synthetic data provided to the system are differentially private, so the privacy of the result depends on the way \qrlew{} rewrites \emph{Reduces} into \emph{DP} equivalent \emph{Relations}.

All \emph{rewriting rules} with \emph{DP} outputs require the input of the \emph{Reduce} to be \emph{PUP} so we can assume a PID column clearly assign one and only one PU to each rows of the rewritten input. The \emph{Reduce} is made DP by:
\begin{itemize}
    \item Making sure the aggregate columns of the \emph{Reduce} are computed with differentially private mechanisms.
    \item Making sure the grouping keys of the \texttt{GROUP BY} clause are either public or released through a differentially private mechanism.
\end{itemize}

\subsubsection{Protecting aggregation results}

The protection of aggregation functions is carried out in two steps. Given that all currently supported aggregations (\texttt{COUNT}, \texttt{SUM}, \texttt{AVG}, \texttt{VARIANCE} \texttt{STDDEV}) can be reduced to sums, our focus will be on \texttt{SUM} aggregations, i.e. the computation of partial sums of a column for different groups: $j\in\{1,\ldots,m\}$, of rows.

Let the column be a vector of $N$ real numbers: $x = \left(x_1,\ldots, x_N\right)\in\mathbb{R}^N$. We note: $\pi_k = i \in \{1,\ldots,n\}$ the PID and $g_k = j \in \{1,\ldots,m\}$ the grouping key associated to $x_k$.
We want to compute all the sums:
$$S_j = \sum_{g_k = j} x_k$$
with some DP guarantees. To this end we:

\begin{enumerate}
    \item \emph{Limit the contribution of each \emph{privacy unit} to the sum}:
    We represent the contribution of each PU: $i$, by a vector: $s_i$ whose components are the partial sums within each of the $m$ groups: $s_i = \left(s_{i,1},\ldots, s_{i,m}\right)$, where:
    $$s_{i,j} = \sum_{\substack{\pi_k = i\\g_k = j}}x_k$$
    The $s_i$'s $\ell^2$ norms are then clipped to $c$:
    $$\overline{s_i} = \left(\overline{s_{i,j}}\right)_j = \left(\frac{s_{i,j}}{\max\left(1, \frac{\|s_i\|_2}{c}\right)}\right)_j$$
    See section~\ref{sec:limit_contrib_per_user} for more details.

    \item \emph{Add gaussian noise to each group}:
    The clipped contributions are summed and perturbed with gaussian noise $\nu = \left(\nu_1,\ldots \nu_m\right) \sim \mathcal{N}\left(0, \sigma^2I_m\right)$:
    $$\widetilde{S_j} = \sum_{i=1}^n \overline{s_{i,j}} + \nu_j$$
    With $\sigma^2={\frac {2\ln(1.25/\delta )\cdot c^{2}}{\varepsilon ^{2}}}$.
    Note that the vector of sums has $\ell^2$ \emph{Global Sensitivity} of $c$, so this is an application of the \emph{Gaussian Mechanism} (see: theorem A.1. in \cite{dwork2014algorithmic}) and the mechanism is $\varepsilon, \delta$-differentially private.
\end{enumerate}

\subsubsection{Protecting grouping keys}

When the grouping keys from a are derived from the data, they are not safe for publication.
Following \cite{korolova2009releasing, wilson2019differentially}, we use a mechanism called \emph{$\tau$-thresholding} to safely release these grouping keys.
Note that, thanks to \emph{range propagation} (see section~\ref{sec:range_propagation}), some groups are already public and need no differentially private mechanism to be published.
Ultimately, the rewriting of: \texttt{SELECT sum(x) FROM table GROUP BY g WHERE g IN (1, 2, 3)} as a DP equivalent will not use \emph{$\tau$-thresholding}, while \texttt{SELECT sum(x) FROM table GROUP BY g} will most certainly do if nothing more is known about \texttt{g} beforehand.

To summarize the various mechanisms used in \qrlew{} to date:
the rewriting of \emph{Reduces} with $PUP \rightarrow DP$ rules requires the use of \emph{gaussian mechanisms} and \emph{$\tau$-thresholding} mechanisms;
then the DP mechanisms used in all the rewritings are aggregated by the \qrlew{} rewriter as a composed mechanism.
The overall privacy loss is aggregated in a RDP accountant \cite{mironov2017renyi}.

\section{Comparison to other systems}

There are a few existing open-source libraries for differential privacy.

Some libraries focus on deep learning and \emph{DP-SGD} \cite{abadi2016deep}, such as: \emph{Opacus} \cite{yousefpour2021opacus}, \emph{Tensorflow Privacy} \cite{TensorFlowPrivacy} or \emph{Optax's DP-SGD} \cite{deepmind2020jax}. \qrlew{} has a very different goal: analytics and SQL.

\emph{GoogleDP} \cite{GoogleDP} is a library implementing many differentially private mechanisms in various languages (C++, Go and Java).
\emph{IBM's diffprivlib} \cite{diffprivlib} is also a rich library implementing a wide variety of DP primitives in python and in particular many DP versions of classical machine learning algorithms. 
These libraries provide the bricks for experts to build DP algorithms. \qrlew{} has a very different approach, it is a high level tool designed to take queries written in SQL by a data practitioner with no expertise in privacy and to rewrite them into DP equivalent able to run on any SQL-enabled data store. \qrlew{} implemented very few DP mechanisms to date, but automated the whole process of rewriting a query, while these library offer a rich variety of DP mechanism, and give full control to the user to use them as they wish.

Google built several higher-level tools on top of \cite{GoogleDP}.
\emph{PrivacyOnBeam} \cite{PrivacyOnBeam} is a framework to run DP jobs written in Apache Beam with its Go SDK.
\emph{PipelineDP} \cite{PipelineDP} is a framework that let analysts write Beam-like or Spark-like programs and have them run on Apache Spark or Apache Beam as back-end. It focuses on the Beam and Spark ecosystem, while \qrlew{} tries to provide an SQL interface to the analyst and runs on SQL-enabled back-ends (including Spark, a variety of data warehouses, and more traditional databases).
\cite{ZetaSQL}, gives the user a way to write SQL-like queries and have them executed on tables using GoogleDB custom code, so it is not  compatible with any SQL data store and support relatively simple queries only.

\emph{OpenDP} \cite{OpenDP} is a powerful Rust library with a python bindings. It offers many possibilities of building complex DP computations by composing basic elements. Nonetheless, it require both expertise in privacy and to learn a new API to describe a query. Also, the computations are handled by the Rust core, so it does not integrate easily with existing data stores and may not scale well either.

\emph{Tumult Analytics} \cite{berghel2022tumult} shares many of the nice composable design of OpenDP, but runs on Apache Spark, making it a scalable alternative to OpenDP. Still, it require the learning of a specific API (close to that of Spark) and cannot leverage any SQL back-end.

\emph{SmartNoise SQL} is a library that share some of the design choices of \qrlew{}. An analyst can write SQL queries, but the scope of possible queries is relatively limited: no \texttt{JOIN}s, no sub-queries, no CTEs (\texttt{WITH}) that \qrlew{} supports. Also, it does not run the full computation in the DB so the integration with existing systems may not be straightforward.

Other systems such as \emph{PINQ} \cite{mcsherry2009privacy} and \emph{Chorus} \cite{johnson2020chorus} are prototypes that do not seem to be actively maintained. \emph{Chorus} shares many of the design goals of \qrlew{}, but requires post-processing outside of the DB, which can make the integration more complex on the data-owner side (as the computation happens in two distinct places).

Beyond that, \qrlew{} brings unique functionalities, such as:
\begin{itemize}
    \item advanced automated range propagation;
    \item the possibility to automatically blend in synthetic data;
    \item advanced privacy unit definition capabilities across many related tables;
    \item the possibility for the non-expert to simply write standard SQL, but for the DP aware analyst to improve its utility by adding \texttt{WHERE x < b} or \texttt{WHERE x IN (1,2,3)} to give hints to the \qrlew{};
    \item all the compute happens in the DB.
\end{itemize}

This last point comes with some limitations (see section~\ref{sec:limitations}), but opens new possibilities like the delegation of the rewriting to a trusted third party. The data practitioner could simply write his desired query in SQL, send it to the rewriter that would keep track of the privacy losses and use \qrlew{} to rewrite the query, sign it, and send it back to the data practitioner that can then send the data-owner, who will check the signature certifying the DP properties of the rewritten query\footnote{A proof of concept is available at: \url{https://github.com/Qrlew/server}}.

\section{Known Limitations}
\label{sec:limitations}

\qrlew{} still implements a limited number of DP mechanisms, it is still lacking basic functionalities such as: quantile estimation, exponential mechanisms.

\qrlew{} relies on the random number generator of the SQL engine used. It is usually not a cryptographic secure random number generator.

\qrlew{} uses the floating-point numbers of the host SQL engine, therefore it is liable to the vulnerabilities described in \cite{casacuberta2022widespread}.

\section{Conclusion}

\qrlew{} is a novel way of bringing DP to analytics. It brings both a unique set of features, an extended coverage of standard SQL, and full execution in the SQL engine, which opens up new ways to integrate a privacy layer in a data practitioner -- data owner relationship.
The code is available on github: \url{https://github.com/Qrlew/qrlew} with a Python bindings: \url{https://github.com/Qrlew/pyqrlew}, a short description: \url{https://qrlew.github.io/}\\
and an interactive demo: \url{https://qrlew.github.io/dp}.

\bibliographystyle{alpha}
\bibliography{qrlew}
\appendix

\section*{Appendix}

\section*{Clipping value used to limit the contribution per user within the aggregations}
\label{sec:limit_contrib_per_user}

In our algorithm, the clipping value $c$ is given by:
\begin{equation}
    c = k \cdot\max ( |\min \textbf{x}|, |\max \textbf{x}|),
\end{equation}
where $\min \textbf{x}$ and $\max \textbf{x}$ are the known bounds of $\textbf{x}$ and $k$ is some parameter of the engine that can be used to trade some lower noise for some extra bias.

\section*{DP evaluation}
We've assessed the differential privacy of our code following the outlined procedure in \cite{wilson2019differentially}. Our tables were constructed using columns of Halton sequences.
Two scenarii were tested:

\begin{itemize}
    \item In the first scenario, each user possessed exactly one row.
    \item In the second scenario, a user could have several rows, with the number of rows owned by one user following a normal distribution centered at half the size of the table.
\end{itemize}

Adjacent databases were created by removing one user compared to the reference database containing all users. Privacy profiles of the underlying distributions were computed using the formula:

\begin{equation}
    \delta(e^{\varepsilon}) = \sup_{k} \sup_{x\in\mathbb{R}} f_{D}(x) - e^{\varepsilon}f_{D_k}(x),
\end{equation}

In this context, $D$ represents the distribution of results when the query is executed on the entire dataset,
and $D_k$ corresponds to the distribution when the query is run on the dataset excluding the data owned by the user $k$.
The algorithm inputting the $(\varepsilon, \delta)$ parameters is indeed $(\varepsilon, \delta)$-differentially private if $\delta(e^{\varepsilon})$ in smaller than $\delta$.
We have verified this property holds true for various queries.

\end{document}